# Inferring incubation period distribution of COVID-19 based on SEAIR Model

Shiyang Lai [a,*], Tianqi Zhao [b], Ningyuan Fan [a]

a. School of Business Administration, Northeastern University, P.R. China.

b. School of Software Engineering, Northeastern University, P.R. China.

**Abstract**:

Objectives:

To reduce the biases of traditional survey-based methods, this paper proposes an epidemic model-based approach to inference the incubation period distribution of COVID-19 utilizing the publicly reported confirmed case number.

Methods:

We construct an epidemic model, namely SEAIR, and take advantage of the dynamic transmission process depicted by SEAIR to estimate the onset probability in each day of exposed individuals in eight impacted countries. Based on these estimations, the general incubation probability distribution of COVID-19 has been revealed.

Results:

The expectation, median, and 95th percentile retrieved from our major result are 6.34(95% credible interval (CI)：6.27-6.42), 4.76(95% CI：4.68-4.84), and 14.12(95% CI：14.04-14.20). We also estimate the incubation period distribution of COVID-19 under different government's containment efforts. The results show that the weaker the control of the government, the closer our predictions are to the results of existing researches.

Conclusion:

The proposed method can avoid several biases of traditional survey-based methods. However, due to the mathematical-model-based nature of this method, the inference results are somewhat sensitive to the setting of parameters. Therefore, this method should be practiced reasonably on the basis of a certain understanding of the studied epidemic.

---

[*] Corresponding author. Email: sl4710@columbia.edu

Declaration of Interest: None

# 1. Introduction

The current outbreak of Coronavirus disease 2019 (COVID-19) has developed into a global pandemic and leads to a global crisis. So far, however, the knowledge of COVID-19 is limited, especially its incubation period distribution. The incubation period of an infectious disease refers to the time interval between infection and onset of symptoms. Understanding the incubation period is not only helpful for disease control and surveillance, but also essential to the study of mechanisms of disease transmission. For instance, the optimal quarantine period can be determined according to the distribution of incubation period. Although the incubation period is of great importance, it is not easy to obtain the accurate distribution of incubation period under data limitation.

The existing literature regarding the estimation of incubation period of COVID-19 mainly relies on sampling and surveys from the infected people. For example, Guan et al., (2020) made a summary statistics of incubation period based on the 291 patients who had claimed a clear specific exposure date. Their results stated that the median incubation period is four days. Backer et al., (2020) estimated the distribution of incubation period using 88 samples that had the travel history from Wuhan, and showed that the Weibull distribution fits the data best and the mean incubation period is 6.4 days (95% credible interval (CI): 5.6–7.7). Qin et al., (2020) utilized the renewal process to estimate the distribution of incubation period using 1211 samples who had clear dates of departure from Wuhan and dates of symptoms onset. They suggested that the median incubation period is 8.13 days (95% CI: 7·37-8·91) and mean incubation period is 8.62 (95% CI: 8.02-9.28).

However, the above-mentioned methods suffer several drawbacks. First, surveys are always along with errors, such as sampling error, coverage error, and measurement error. The sample size of current research ranges from 88 to 1211. Under the limitation of sample size, it is challenging to make a reliable estimation of incubation period distribution. Second, when surveying, the patients' recall bias is inevitable, which may also lead to erroneous results. Some patients may not know the exact exposure date, or maybe lack of the memory of exposure history. Therefore, the exposure dates may not be accurately monitored and recorded. Last but not least, the judgment of contact trackers will also influence the determination and record of the exposure date, thus influencing the accuracy of the estimation of incubation period. To address the above shortcomings, we draw on the idea of the SEAIR model and propose a novel method to infer the probability distribution of the incubation period of COVID-19 based on the daily confirmed case number. Contrary to the traditional survey method, the proposed method reduces the dependence on survey data.

# 2. Method

Since the simple SIR model was proposed by William et al. in 1927, a large number of advanced epidemiological models have been generated on this basis to explain different kinds of real word infectious disease transmission (Calatayud et al., 2018; Chinazzi et al., 2020; Mizumoto et al., 2020; Arino & Portet, 2020). In the specific context of COVID-19, we proposed a SEAIR epidemic model with a time delay that follows a discrete random probability distribution (Figure 1).

*Figure 1 here*

In this model, we define the total population of the target country as *N*, and all the people in this country can be divided into five epidemiological classes, *S*, *E*, *A*, *I*, and *R*, respectively. We call $S_t$ the number of susceptible individuals in the country at time *t*, $E_t$ the number of exposed individuals at time *t*, $I_t$ the number of infected people with symptoms at time *t*, $A_t$ the number of infected people who are asymptomatic at time *t*, and $R_t$ the number of removed individuals including recovered and dead cases at time *t*. Then, we have $N = S_t + E_t + A_t + I_t + R_t$. In order to model discrete incubation periods, the following assumptions are established based on several previous studies about COVID-19:

(**A1**) The whole population of the target country *N* is constant over time since the international transportation of most countries are strictly restricted after the outbreak of COVID-19 (Chinazzi et al., 2020; Salcedo et al., 2020).

(**A2**) There is a damping effect on the virus reproduction rate due to the government's containment effort, and this effect can be modeled through an exponentially decreasing function (Lanteri et al., 2020). Namely, $\alpha = R0 \times e^{-\lambda t}$

(**A3**) Asymptomatic individuals are hard to be recorded, and their number accounts for a certain ratio of all infected people (Mizumoto et al., 2020). Namely, $A_t = \gamma I_t$

(**A4**) Once the infected people develop symptoms, they will be recorded and quarantined just on the day of the onset. Therefore, they will lose the ability to spread the virus after this day. On the contrary, asymptomatic individuals can infect others continually since they are difficult to identify (Li et al., 2020). Namely, $\Delta E_t = S_t \times \dfrac{A_t + \Delta I_t}{N} \times \alpha$

Based on the above-mentioned assumptions, we try to describe the dynamic process from group *E* to group *A* and *I*, including a discrete-time delay. The newly added symptomatic and asymptomatic cases' number can be interpreted as the number of all exposed individuals who were infected before day *t* and become sick on day *t*. Through this logic, we can define the sum of $\Delta I_t$ and $\Delta A_t$ as follow:

$$\Delta I_t + \Delta A_t = \Delta E_{t-k} \times p_k + \Delta E_{t-k+1} \times p_{k-1} + \cdots + \Delta E_{t-1} \times p_1 + \epsilon_t \tag{1}$$

In this equation, $P_k$ is the probability of the exposed individual having a $k$-day incubation period. Also, because of the assumption A3, we have $\Delta A_t = \gamma \Delta I_t$. Therefore

$$\Delta I_t = \frac{\Delta E_{t-k} \times p_k + \Delta E_{t-k+1} \times p_{k-1} + \cdots + \Delta E_{t-1} \times p_1}{1+\gamma} + \epsilon_t \tag{2}$$

To make this equation calculable, we set a threshold $k$, which represents the maximum length of the incubation periods that this epidemic could have. In other words, the probability of exposed individuals generated before day $t-k$ falling ill on day $t$ is zero.

In a $T$ days long time window where $T > k$, this problem can be transformed into a problem of solving multivariate linear equations. Given a $T$ length time series, we define

$$\Delta I = (\Delta I_{k+1}, \Delta I_{k+2}, \cdots, \Delta I_T)^\top, \tag{3}$$

$$\Delta E = \begin{pmatrix} \Delta E_1 & \Delta E_2 & \cdots & \Delta E_k \\ \Delta E_2 & \Delta E_3 & \cdots & \Delta E_{k+1} \\ \vdots & \vdots & \ddots & \vdots \\ \Delta E_{T-k-1} & \Delta E_{T-k} & \cdots & \Delta E_{T-1} \end{pmatrix}, \tag{4}$$

$$\mathbf{p} = (p_k, p_{k-1}, \cdots, p_1)^\top. \tag{5}$$

Note that $\forall p_i \in \mathbf{P}, p_i \geq 0$, and $\sum p_i = 1$.

We can estimate $P$ through minimizing the least square cost function of the multivariate linear equations presented below:

$$\arg\lim_{\mathbf{p}} \sum_i \left( \Delta I - \Delta \hat{I} \right)^2, \tag{6}$$

where $\Delta \hat{I} = \frac{\Delta E \times \mathbf{p}}{1+\gamma}$.

However, due to the high-dimensional characteristic of this objective function, this task is likely to have a lot of local minima which makes it hard for standard optimization methods because there is a strong dependency on the initial condition. We employ the basin-hopping algorithm, a global optimization method introduced by Wales et al. in 1998 to obtain the estimation of $P$.

## 3. Results

We retrieve the publicly available data of the COVID-19's confirmed, recovered, and death cases from ministries of health in eight countries from January 21 to April 13. Yang et al. (2020) estimate that the reproduction rate of COVID-19 is 3.77, using 8,866 cases recorded in China. Referring to their work, the value of *R0* in our model is also set to 3.77. Meanwhile, based on recent research about the Diamond

Princess ship event (Mizumoto et al., 2020), the estimated asymptomatic proportion among all infected cases is about 17.9%. Thus, we value $\gamma$ to 22% (The ratio of asymptomatic patients to symptomatic patients is 1 to 4.56). Besides, since several epidemiological studies have proven that the incubation period of COVID-19 is hardly longer than 20 days (99th percentile) (Jing et al., 2020; Backer et al., 2020), so we set the threshold $k$ to 20. We use the historical data of one month after the outbreak of each country to calculate $P$ and use spline to construct the probability distribution of the incubation period. Figure 2 shows the estimated incubation period distribution when $\lambda$ is set to 0.1. The subgraph on the left side presents the results of each country and the overall mean value of these countries, and the child plot in the top right corner presents the changing of reproduction number when $\lambda$ equals 0.1. The shade part of the subgraph on the right side is the 95% pointwise confidence interval (CI) of the estimated distribution.

*Figure 2 here*

As we can see in Figure 2, basically, among all the studied countries, the inference results are similar. In this estimated incubation period probability distribution, the distribution reaches its peak on the fourth day, meaning a person infected with COVID-19 is most likely to have a four-day incubation period. However, in the real-world situation, we cannot identify the 'real' $\lambda$, which indicates the effect of the countries' measures on suppressing the spread of epidemics. Accordingly, we also operate sensitivity analysis to $\lambda$ to test the stability of our estimation, and the results have been shown in Figure 3.

*Figure 3 here*

With five different values of $\lambda$, our predicted distribution still maintains a similar shape and consistent peak position. But as $\lambda$ becomes larger, the distribution becomes more and more smooth. We list the expectation, the median, and the 95th percentile under these five different conditions in table 1.

*Table 1 here*

Comparing with several current studies about COVID-19's incubation period (Li et al., 2020; Guan et al., 2020; Backer et al., 2020; Linton et al., 2020), our listed inferences' results are quite close to theirs, especially when $\lambda$ is equal to 0.01 or 0.05. This phenomenon might imply the fact that, after the outbreak of COVID-19, most countries did not operate efficiently and urgently to suppress the spreading of the virus.

## 4. Discussion

In this paper, we introduce an epidemic-model-based approach to inference the distribution of the incubation period of the COVID-19, and our results are in line with the existing studies. The contributions of this paper are twofold. First, based on the transmission process depicted by the established epidemic model, our proposed method infers the distribution of incubation period from the overall daily confirmed cases number reported by each countries' governments. Therefore, our method avoids the drawbacks associated with traditional survey-based methods. Second, distinct with the traditional methods, our approach doesn't need to make prior assumptions about the distribution of the incubation periods. We provide a more actual estimation result that is closer to the "real" situation to some extent. Nevertheless, because the proposed approach is based on the prior knowledge of COVID-19, such as reproduction rate, asymptomatic proportion, government's containment effort, and so on, which is needed to depict the transmission process, our results are sensitive to the setting of parameters. Thus, to make reliable inference using this method, information about the virus provided by prior researches are necessary.

# References

Arino, J., & Portet, S. (2020). A simple model for COVID-19. *Infectious Disease Modelling, 5,* 309-315. DOI: 10.1016/j.idm.2020.04.002

Backer, J. A., Klinkenberg, D., & Wallinga, J. (2020). Incubation period of 2019 novel coronavirus (2019-nCoV) infections among travellers from Wuhan, China, 20–28 January 2020. *Eurosurveillance, 25*(5). DOI: 10.2807/1560-7917.es.2020.25.5.2000062

Calatayud, J., Cortés, J. C., Jornet, M., & Villanueva, R. J. (2018). Computational uncertainty quantification for random time-discrete epidemiological models using adaptive gPC. *Mathematical Methods in the Applied Sciences, 41*(18), 9618–9627. DOI: 10.1002/mma.5315

Chinazzi, M., Davis, J. T., Ajelli, M., Gioannini, C., Litvinova, M., Merler, S., … Vespignani, A. (2020). The effect of travel restrictions on the spread of the 2019 novel coronavirus (COVID-19) outbreak. *Science*. DOI: 10.1126/science.aba9757

Guan, Wei-Jie., Ni, Zheng-yi & Hu, Yu et al. (2020). Clinical Characteristics of Coronavirus Disease 2019 in China. *New England Journal of Medicine*. DOI: 10.1056/NEJMoa2002032

Lanteri, D., Carco', D., & Castorina, P. (2020). *How macroscopic laws describe complex dynamics: asymptomatic population and CoViD-19 spreading*. Retrieved 25 April 2020, from https://arxiv.org/abs/2003.12457

Li, Qun & Guan, Xuhua & Wu, Peng & Wang, Xiaoye & Zhou, Lei & Tong, Yeqing & Ren, Ruiqi & Leung, Kathy & Lau, Eric & Wong, Jessica Y & Xing, Xuesen & Xiang, Nijuan & Wu, Yang & Li, Chao & Chen, Qi & Li, Dan & Liu, Tian & Zhao, Jing & Li, Man & Feng, Zijian. (2020). Early Transmission Dynamics in Wuhan, China, of Novel Coronavirus–Infected Pneumonia. *New England Journal of Medicine*. DOI: 382. 10.1056/NEJMoa2001316

Li, R., Pei, S., Chen, B., Song, Y., Zhang, T., Yang, W., & Shaman, J. (2020). Substantial undocumented infection facilitates the rapid dissemination of novel coronavirus (SARS-CoV-2). *Science, 368*(6490), 489–493. DOI: 10.1126/science.abb3221

Linton, Natalie & Kobayashi, Tetsuro & Yang, Yichi & Hayashi, Katsuma & Akhmetzhanov, Andrei & Jung, Sung-Mok & Yuan, Baoyin & Kinoshita, Ryo & Nishiura, Hiroshi. (2020). Incubation Period and Other Epidemiological Characteristics of 2019 Novel Coronavirus Infections with Right Truncation: A Statistical Analysis of Publicly Available Case Data. *Journal of clinical medicine. 9.* DOI: 10.3390/jcm9020538

Mizumoto, Kenji & Kagaya, Katsushi & Zarebski, Alexander & Chowell, Gerardo. (2020). Estimating the asymptomatic proportion of coronavirus disease 2019 (COVID-19) cases on board the Diamond Princess cruise ship, Yokohama, Japan, 2020. *Eurosurveillance*. 25. DOI: 10.2807/1560-7917.ES.2020.25.10.2000180

Jing, Q., You, C., Lin, Q., Hu, T., Yu, S., & Zhou, X.-H. (2020). Estimation of incubation period distribution of COVID-19 using disease onset forward time: a novel cross-sectional and forward follow-up study. *medRxiv*. DOI: 10.1101/2020.03.06.20032417

Salcedo, A., Yar, S., & Cherelus, G. (2020). *Coronavirus Travel Restrictions, Across the Globe.* Retrieved May 3, 2020, from https://www.nytimes.com/article/coronavirus-travel-restrictions.html

Wales, David & Doye, Jonathan. (1998). Global Optimization by Basin-Hopping and the Lowest Energy Structures of Lennard-Jones Clusters Containing up to 110 Atoms. *The Journal of Physical Chemistry A*. DOI: 101. 10.1021/jp970984n

William Ogilvy Kermack, A. G. McKendrick and Gilbert Thomas Walker. (1927). A contribution to the mathematical theory of epidemics. *Proc. R. Soc. Lond*. A115700–721. http://doi.org/10.1098/rspa.1927.0118

Yang Yang, Qingbin Lu, Mingjin Liu, Yixing Wang, Anran Zhang, Neda Jalali, Natalie Dean, Ira Longini, M. Elizabeth Halloran, Bo Xu, Xiaoai Zhang, Liping Wang, Wei Liu, Liqun Fang. (2020). Epidemiological and clinical features of the 2019 novel coronavirus outbreak in China. *medRxiv*. DOI: https://doi.org/10.1101/2020.02.10.20021675

**Table 1.** Expectation and 95th percentile in different $\lambda$

|  | $\lambda = 0.01$ | $\lambda = 0.05$ | $\lambda = 0.10$ | $\lambda = 0.15$ | $\lambda = 0.20$ |
|---|---|---|---|---|---|
| Expectation (95%CI) | 5.16 (5.07-5.25) | 5.50 (5.41-5.58) | 6.34 (6.27-6.42) | 7.24 (7.18-7.31) | 6.76 (6.71-6.81) |
| Median (95%CI) | 3.54 (3.45-3.63) | 3.91 (3.83-3.99) | 4.76 (4.68-4.84) | 5.74 (5.67-5.81) | 3.93 (3.88-3.98) |
| 95th percentile (95%CI) | 13.85 (13.76-13.94) | 12.86 (12.78-12.94) | 14.12 (14.04-14.20) | 15.76 (15.69-15.83) | 15.82 (15.77-15.87) |

**Figure 1.** Flowchart of an SEAIR epidemic model that has a time delay with a discrete random probability distribution

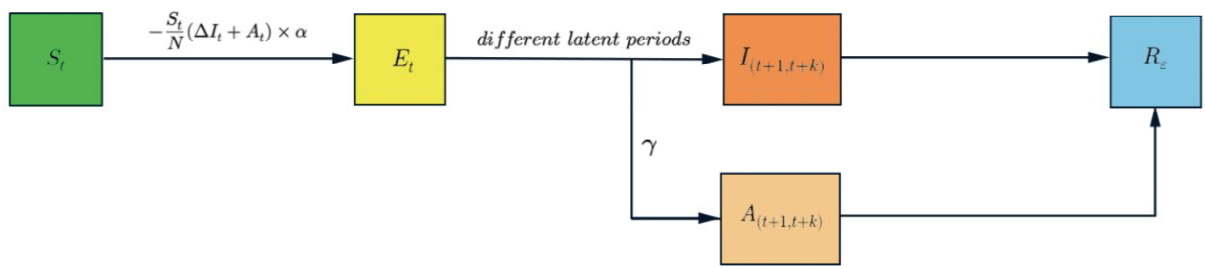

**Figure 2.** Inference the probability distribution of incubation period when $\lambda = 0.1$

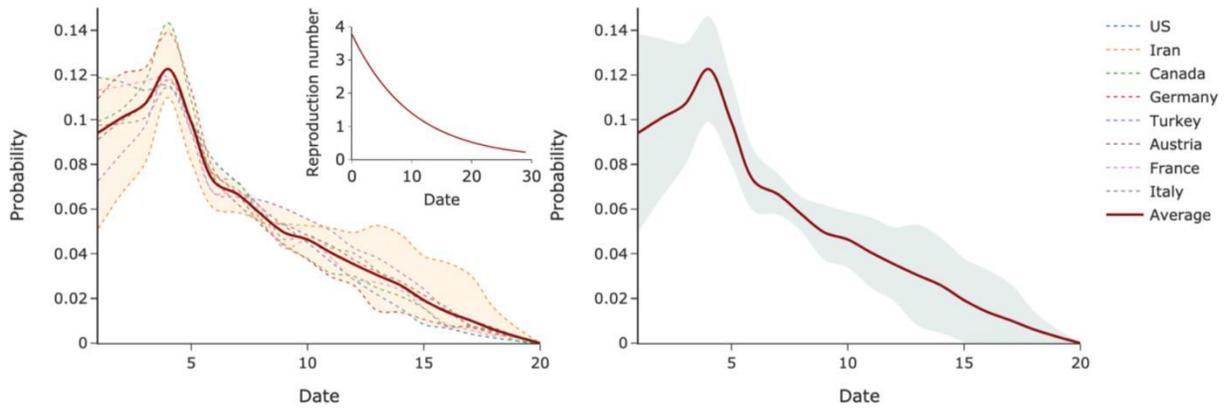

**Figure 3.** Inference the probability distribution of incubation period in different $\lambda$. (The $\lambda$ in the upper left subgraph is 0.01, in the upper right is 0.05, in the lower left is 0.15, and in the lower right is 0.20. The child graph in each subgraph's upper right corner shows the change of reproduction number with date under the corresponding $\lambda$)

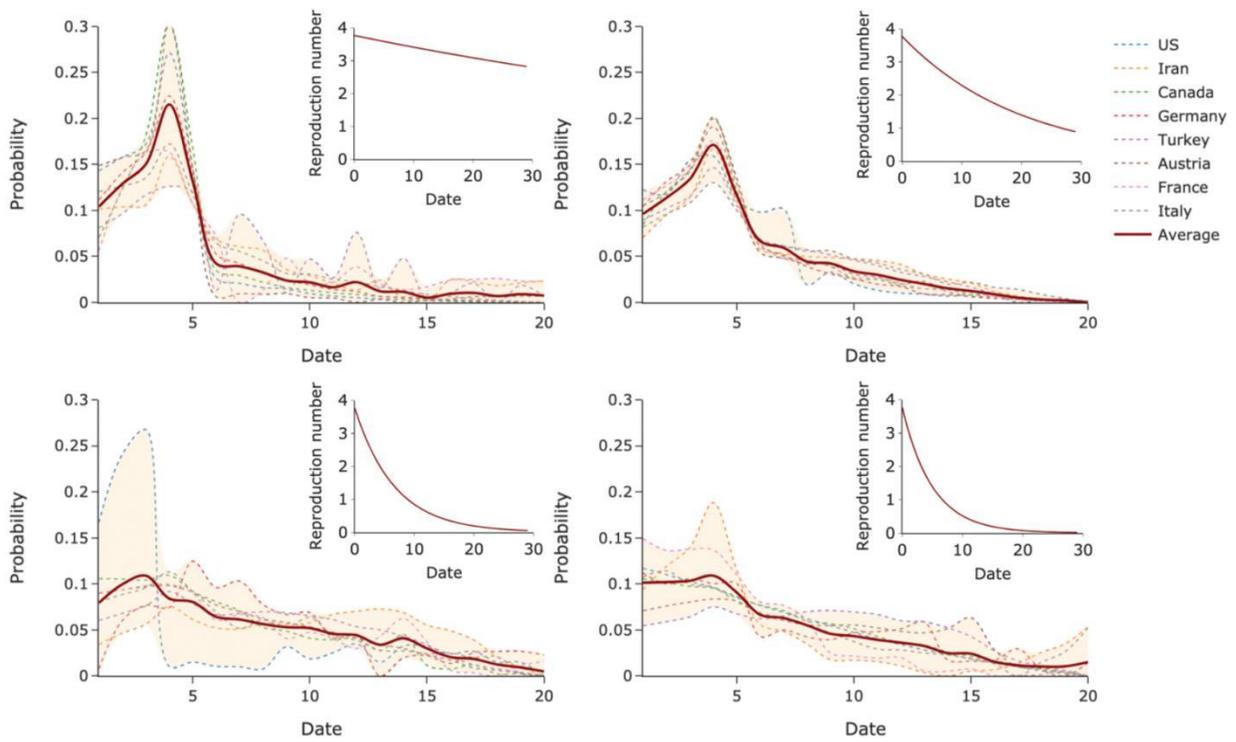